\documentclass[aps,pre,floats,floatfix,twocolumn,longbibliography,superscriptaddress]{revtex4-2}

\usepackage{color} 

\usepackage{newtxmath}

\usepackage{latexsym}
\usepackage{amsmath} 
\usepackage{bm}
\usepackage{times}
\usepackage{soul}
\usepackage{float}

\usepackage{graphicx}
\usepackage{epstopdf}
\graphicspath{{figs/}}

\usepackage[colorlinks,linkcolor=blue,urlcolor=blue,citecolor=blue,bookmarks=true,pdftitle={}]{hyperref}

\DeclareSymbolFont{myletters}{OML}{ztmcm}{m}{it}
\DeclareMathSymbol{\uplambda}{\mathord}{myletters}{"15}

\DeclareSymbolFont{cmlargesymbols}{OMX}{cmex}{m}{n}
\let\sum\relax
\DeclareMathSymbol{\sum}{\mathop}{cmlargesymbols}{"50}
\DeclareSymbolFont{cmletters}{OML}{cmm}{m}{it}
\SetSymbolFont{cmletters}{bold}{OML}{cmm}{b}{it}
\DeclareSymbolFontAlphabet{\mathnormal}{cmletters}

\newcommand{\bra}[1]{\left\langle #1 \right|}
\newcommand{\ket}[1]{\left| #1 \right\rangle}

\newcommand{\beq}{\begin{eqnarray}}
\newcommand{\eeq}{\end{eqnarray}} 

\newcommand{\hide}[1]{}

\begin{document}

\title{Supersolid phases of bosons}
 
\author{Sudip Sinha}
\email[]{sudip.iiserk2016@gmail.com}
\affiliation{Indian Institute of Science Education and Research Kolkata, Mohanpur, Nadia 741246, India}

\author{Subhasis Sinha}
\affiliation{Indian Institute of Science Education and Research Kolkata, Mohanpur, Nadia 741246, India}
 
\date{\today}

\begin{abstract}
Supersolids—the enigmatic phase of quantum matter, with properties resembling both the superfluid and solid states—have been actively sought over the past 70 years. We provide a comprehensive review of the developments to date in experimental and theoretical studies of the supersolid phases of bosons, with a particular focus on their observation in ultracold atomic gases. Additionally, the use of optical lattices facilitates the realization of `lattice-supersolids', which paves the way to study the effect of correlations in a controlled manner. A brief theoretical framework is presented to characterize this puzzling state with competing orders and to gain insight into its basic properties. Various types of supersolid phases and the different platforms used to achieve them are described. Finally, we discuss the future prospects of research and the potential to achieve supersolids with more exotic features.
\end{abstract}

\maketitle

\section{Introduction: what is a supersolid?}
The search for novel and exotic phases of quantum matter with coexisting order is a subject of intense research. 
The discovery of superfluidity in Helium ($^4$He) \cite{Kapitza1938, Misener1938} sparked a profound quest to search for the enigmatic `superfluid-solid' or supersolid (SS) phase, where both the superfluid (SF) and solid order can exist simultaneously \cite{Prokofev_review2007, Balibar_review2008, Balibar_review2010, Prokofev_review2012, Reatto_review2013, Yukalov_review2020, Bottcher_review2021, Stringari_review2023}.
Superfluidity is related to the dissipationless flow, which is a purely quantum mechanical phenomenon that can be mathematically quantified from the `off-diagonal long-range order' (ODLRO) \cite{Yang1962}.
On the contrary, the structural ordering of particles in a solid can be characterized by `diagonal long-range order' (DLRO). 
In this context, supersolidity can be described by the emergence of both DLRO and ODLRO simultaneously. 
In case of bosons, the ODLRO is associated with the phenomenon of Bose-Einstein condensation (BEC) \cite{Penrose1951,Penrose_and_Onsager1956}, where a fraction of particles occupy the lowest energy state at extremely low temperature. 
The question related to the existence of the ODLRO in solids was first raised by Penrose and Onsager \cite{Penrose_and_Onsager1956} and was later revisited in a series of works \cite{Andreev1969, Thouless1969, Chester1970, Leggett1970, Matsuda1970, Guyer1971, Mullin1971, Imry1975, Andreev1982_review, Nozieres1995}.
The possibility of BEC associated with ODLRO in a quantum crystal was discussed by Chester \cite{Chester1970} and finding the signature of supersolidity from the non-classical rotation inertia was proposed by Leggett \cite{Leggett1970}. 
The SS state is counterintuitive since a solid is described by localized particles, whereas superfluidity is attributed to the dissipationless flow of collective matter waves with phase coherence.
Moreover, this state is paradoxical in the sense that the liquid-like superflow of the SF coexists with the elastic shear stiffness similar to solids.
In this context, another route to achieving supersolidity was proposed by Andreev and Lifshitz \cite{Andreev1969}, where zero-point crystal defects can undergo BEC at low temperature, forming an SF within the crystal. 
This issue was also independently addressed by Thouless \cite{Thouless1969} and later explored by others \cite{Chester1970, Guyer1971, Mullin1971, Imry1975, Prokofev_Svistunov2005}.
However, Anderson proposed a different view regarding the supersolidity of $^4$He \cite{Anderson_Nature2007,Anderson2008,Anderson2009}.
Starting from the lattice gas model \cite{Matsuda1970}, such an intriguing phase has been explored theoretically in various bosonic and spin models (see for example Ref.~\cite{Prokofev_review2012} and references therein).  
Although the idea of supersolidity is originally associated with the phases of $^4$He, its existence in this system remains elusive.
The experiment by Kim and Chan \cite{Kim_nature2004,Kim_science2004} stimulated the search for the SS phase of $^4$He \cite{Sasaki2006, Reppy2006, Reppy2007, Kondo2007, Kojima2007, Clark2007, Kubota2007, Kubota2008, Hallock2008, Balatsky2009, Choi2010, Choi2010_2, Toda2010, Kim_NJP2010, Golov2011, Balibar2012, Reppy2016}, as they observed a drop in the rotational inertia.
However, this drop is also associated with an increase in shear modulus (stiffening) of solid $^4$He \cite{Beamish_review2020, Beamish_Day2007, Beamish_Chan2009, Beamish_day2009, Beamish2012} and from further analysis, it was concluded that the existence of the SS phase is yet to be confirmed \cite{Kim_PRL2012,Kim_PRB2014,Voss2012}.  

\begin{figure*}
	\centering
	\includegraphics[width=0.95\textwidth]{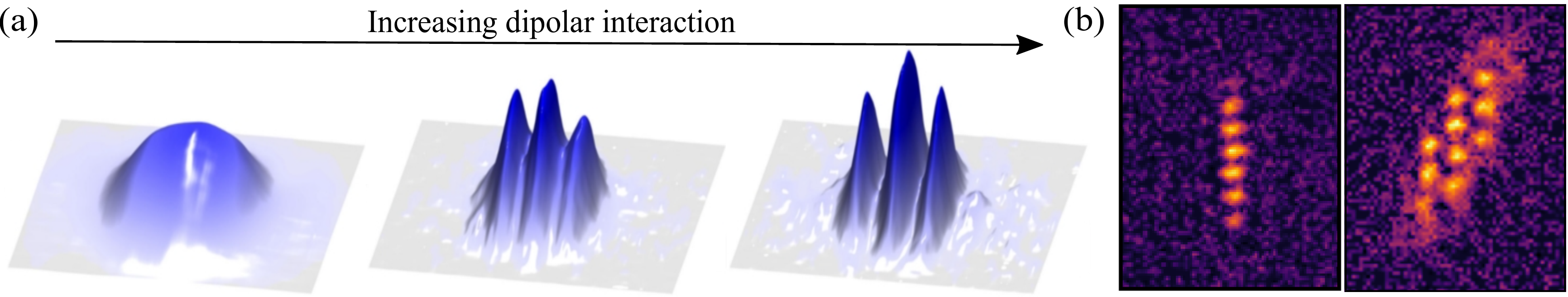}
	\caption{Supersolids of dipolar gas: experimental images of three possible phases of a dipolar gas with increasing dipolar interaction strength, (left) a superfluid (SF) Bose-Einstein condensate (BEC), (middle) the supersolid (SS) phase, and (right) a droplet crystal. 
Within a certain range of parameter, the SS phase features the solid (density modulation) and SF order simultaneously, as an array of droplets is immersed in a condensate background. Reproduced from Ref.~\cite{Bottcher_review2021}. $\copyright$ IOP Publishing Ltd. All rights reserved.
(b) Single-trial images of density profiles of (left) 1D and (right) 2D SSs created with dipolar $^{164}$Dy atoms. The color-scale represents the density from low (black) to high (yellow). Adapted from Ref.~\cite{Ferlaino_exptnature_SS2021}, with permission from Springer nature.
}      
\label{Fig_supersolid_dipolar_gas}
\end{figure*}

Following the unsuccessful attempts to observe supersolidity in $^4$He \cite{Meisel1992, Andreev_expt1969, Suzuki1973, Tsymbalenko1976, Greywall1977, Bishop1981, Dyumin1989, Goodkind1990, Haar1992}, the focus shifted towards ultracold atomic systems, where tremendous experimental advancements facilitated the exploration of various exotic quantum phases \cite{Bloch2005, Sanpera2007, Lewenstein_book2012, Dalibard2008, Bloch2008, Bloch2012, Langen2015, Gross2017, Sugawa2020}. 
Unlike $^4$He, nearly pure Bose-Einstein condensation has been observed in dilute atomic vapors \cite{Cornell1995, Ketterle1995, Ketterle1996, Dalfovo1999, Leggett2001, Grimm2003, Grimm2009}, where the collective matter wave can be represented in terms of classical fields that obey non-linear equations of motion, also known as Gross-Pitaevskii equations (GPE) \cite{Gross1957, Gross1958, Gross1961, Gross1963, Pitaevskii1961}. Historically, shortly after the debate on supersolidity began, it was shown that these non-linear equations admit a periodically modulated ground state solution for certain two-body interactions \cite{Gross1957,Gross1958}. Moreover, such periodic solutions can also arise for a moving condensate \cite{Pitaevskii1984, Pitaevskii2005, Baym2012, Pavloff2021}. 
It turns out that the long-range interaction is a key ingredient for the appearance of density modulations \cite{Gross1958,Fisher1973}; however, the condensates of most of the atomic species are dominated by short-range two-body interactions. As a result, the absence of the long-range interactions in such cases hinders the formation of modulated phases as ground states.
This problem was circumvented after achieving the condensates with dipolar interaction \cite{Lewenstein_review2009, Zoller_review2012, Ferlaino_review2023}, such as Chromium (Cr) \cite{Pfau2005_Chromium}, Erbium (Er) \cite{Ferlaino2012_Erbium}, and Dysprosium (Dy) \cite{Lev2011_Dysprosium} atoms. Moreover, the polar molecules also serve as another promising candidate due to their electric dipole moments \cite{Buchler2007, Ni2008, Rey2013, Rey_review2017, Ye_review2017, BEC_polar2024}.
Despite the long-range nature, the anisotropy of the dipolar interactions can generate instability, leading to the collapse of the dipolar condensates. Nonetheless, recent theoretical studies indicate that the quantum fluctuations (Lee-Huang-Yang (LHY) corrections \cite{LHY_original_paper1,LHY_original_paper2}) can prevent such collapse and lead to the formation of self-bound quantum droplets \cite{Bottcher_review2021, Petrov2015, Tarruell_expt2018, Modugno_expt2018, Pfau_expt_2016, Pfau_expt2_2016, Pfau_exptPRL_2016, Ferlaino_expt_2016, Pfau_droplet_review2018, Pfau_droplet_review2021, Liu_droplet_review2021, Malomed_droplet_review2021}. 
Remarkably, in a narrow region of parameter space, an array of phase-coherent dipolar droplets has been observed experimentally, which in turn leads to the formation of a droplet SS phase \cite{Bottcher_review2021, Pfau_expt_SS2019, Modugno_expt_SS2019, Ferlaino_expt_SS2019, Ferlaino_expt_SS2021, Ferlaino_exptnature_SS2021, Ferlaino_expt_SS2022, Giamarchi_expt_SS2021}. 
As illustrated in Fig.~\ref{Fig_supersolid_dipolar_gas}(a), this phase appears in an intermediate regime between the SF and the droplet crystal, and its density modulation associated with supersolidity has also been observed experimentally [see for example Fig.~\ref{Fig_supersolid_dipolar_gas}(b)].
It should be noted that, the appearance of such density modulated structure breaks the continuous translation symmetry in homogeneous systems. 
As a consequence of this spontaneous symmetry breaking, phonon-like gapless {\it Goldstone} modes appear, which serve as a signature of such SS phase \cite{Modugno_expt_gapless_mode_SS2019, Mossman_review2019, Pfau_expt_gapless_mode_SS2019, Ferlaino_expt_gapless_mode_SS2019}.
Moreover, in the case of SSs, a solid-like elastic response is expected to be accompanied with superfluidity. For a continuum system, such elastic properties can be derived using GPE to obtain the hydrodynamic description of sound modes \cite{Pomeau_Rica2007_1, Dorsey2010, Stringari_sound_2024, Blakie_exc_spec_2024, Blakie_exc_spec2_2024, Rakic2024, Blakie2025, Zwerger2021}. It has been shown that, a defect-free SS can also lower the rotational inertia due to solid-like response \cite{Pomeau_Rica2007_2}.
Apart from dipolar gases, Rydberg atoms \cite{Bloch_expt_Rydberg2012, Pfau_Rydberg_review2012, Lahaye_review2020, Tian_review2021} also serve as another promising candidate for the SS phase \cite{Cinti2010, Pohl_Rydberg2010, Pohl_Rydberg2012, chiral_SS_SOC1, Lyu2020, Wang2022, Eggert2013, Kush2014, Hofstetter2017, Hofstetter2018, Hofstetter2018_quasiparticle_spectraSS, Hofstetter2019_non_eqSS, Hofstetter2019, Hofstetter2022, Pollet_Rydberg2025, chiral_SS_SOC2}, owing to the long-range van der Waals interaction. 
While the inherent dissipation due to spontaneous decay of Rydberg excited states may limit their lifetime, the SS phase can still be achieved as a non-equilibrium state \cite{Hofstetter2019_non_eqSS}.
Moreover, a super-stripe phase sharing the similar properties of SS has also been achieved by engineering a synthetic spin-orbit coupled BEC \cite{Spielman2011, Ketterle2017, Leticia_expt_2024, SOC_BEC2011, Spielman_SOC_review, Stringari_SOC_BEC_review2015, SOC_BEC_review2015, SOC_BEC_review2016}.
Another route to density ordering has recently been engineered by coupling a BEC with cavity modes \cite{Esslinger_expt_2007, Cavity_QED_review}, which induces infinite-range interaction, leading to the formation of SS structure in the condensate \cite{Esslinger_expt_2010, Esslinger_expt_2010, Hemmerich_expt_2015, Esslinger_expt_2017, Esslinger_expt2_2017, Zimmerman_expt_2020}.

Unlike the dilute gases, the effect of correlations in the quantum phases can be studied by loading the cold atoms in an optical lattice, where interactions play a significant role \cite{Bloch2005, Sanpera2007, Lewenstein_book2012, Bloch2008, Gross2017, Sugawa2020, Zoller1998, Zoller2016}.
For example, the seminal experiment of SF to Mott insulator (MI) transition of cold atoms in an optical lattice \cite{Greiner2002} opens up new avenues to simulate strongly correlated exotic phases.
Solid phases with different density ordering have already been achieved in a recent experiment with dipolar atoms in an optical lattice \cite{Griener2023}. 
In this context, the magnetic ordered phases similar to solids have also been emulated in a Rydberg dressed spin lattice \cite{Pohl_Rydberg2016}.
In a remarkable experiment, the cavity-mediated long-range interaction has been employed to obtain the `lattice-SS' phase of bosons in an optical lattice \cite{Esslinger_expt_2016}, as shown schematically in Fig.~\ref{Fig_BHM_schematic}. 
In the case of lattice-SSs, the discrete lattice translational symmetry is broken in contrast to the spontaneous breaking of continuous translational symmetry in homogeneous systems. The discrete $\mathbb{Z}_{2}$ symmetry breaking is associated with the formation of a self-organized lattice-SS phase of BEC coupled to a single cavity, which has been explored experimentally \cite{Esslinger_expt_2010, Hemmerich_expt_2015}.
Such platforms in optical lattice offer an opportunity to study the competing orders in the presence of correlations and show potential for emulation of more exotic phases, which are rare and challenging to observe in conventional materials. 

The purpose of this review is twofold, (i) to offer a brief overview of the theoretical and experimental progress for the search of the SS phase, particularly focusing on their recent observations in ultracold atomic setups; (ii) to provide a theoretical background for characterizing the SSs, as well as highlight the different routes and platforms for achieving supersolidity.

The review is organized as follows. In Sec.~\ref{BHM_and_its_extension}, we introduce the extended Bose-Hubbard model and discuss the different phases it exhibits. Next, we discuss the density ordering of SSs and the effect of thermal fluctuations in Sec.~\ref{density_orderings_in_supersolids}. In Sec.~\ref{Different_directions_for_achieving_supersolidity}, the different directions for achieving supersolidity are listed and elaborated. The signature of SSs from collective excitations is discussed in Sec.~\ref{signature_of_supersolids_from_collective_excitations}. Finally, we conclude with discussion and outlook in Sec.~\ref{discussion_and_outlook}.

\section{Bose-Hubbard model and its extension}
\label{BHM_and_its_extension}
Originally proposed to describe the properties of interacting electrons in solids, the Hubbard model and its variants now serve as cornerstones for understanding the physics of strong correlations \cite{Lewenstein_book2012}. In particular, following the advancement of cold atom experiments, the Hubbard model has been extended to describe the behavior of bosons loaded in optical lattices. 
Within the tight binding approximation, bosons in a lattice can be described by the Bose-Hubbard model (BHM),
\begin{equation}
\hat{\mathcal{H}}_{\text{BHM}} = -\sum^{\text{NN}}_{\langle i,j \rangle}t(\hat{a}^\dagger_{i}\hat{a}_{j} + {\rm h.c}) + \frac{U}{2}\sum_{i}\hat{n}_{i}(\hat{n}_{i}-1) - \mu \hat{n}_{i}, \label{BHM}
\end{equation}
where $\hat{a}^\dagger_{i}$ ($\hat{a}_{i}$) denotes the creation (annihilation) operator of bosons at the $i^{\rm th}$ site of the lattice, and $\hat{n}_{i}=\hat{a}^\dagger_{i}\hat{a}_{i}$ represents the boson number operator. The first term describes the nearest-neighbor (NN) hopping of bosons with hopping amplitude $t$, the second term accounts for the on-site interaction between bosons with strength $U$, and the last term represents the energy required to add a particle at a site, where $\mu$ is the chemical potential. This paradigmatic model captures the transition from MI to SF phase of bosons \cite{Fisher1989}, which has already been observed in a seminal experiment \cite{Greiner2002}. Both phases are homogeneous and preserve the lattice translation symmetry. 

\begin{figure}
	\centering
	\includegraphics[width=\columnwidth]{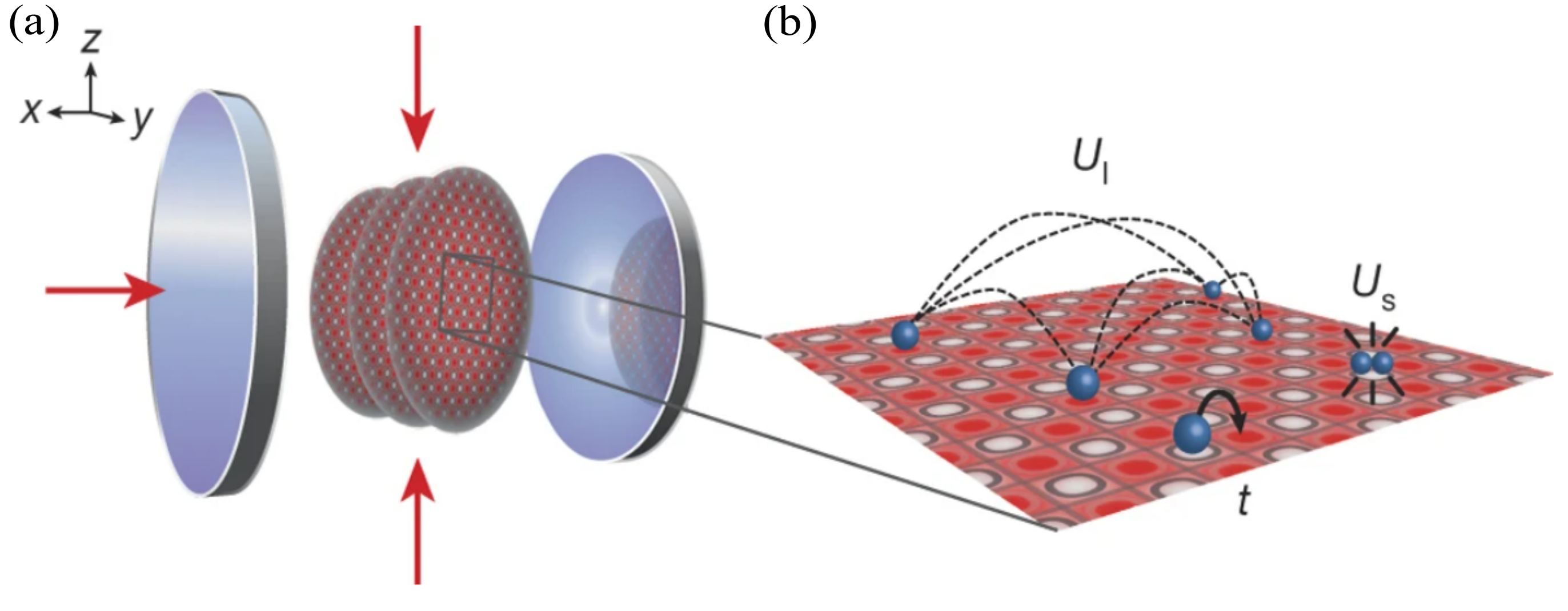}
	\caption{ Illustration of the experimental scheme that realizes a Extended Bose-Hubbard model with on-site and infinite-range interactions.
(a) Stack of 2D systems along the y axis loaded into a 2D optical lattice (red arrows) between two mirrors (shown grey). The cavity induces atom–atom interactions of infinite-range. (b) Illustration of the competing energy scales: tunnelling $t$, on-site interactions $U_{s}$ and long-range interactions $U_{l}$. Figure adapted from Ref.~\cite{Esslinger_expt_2016}, with permission taken from Springer Nature.}     
\label{Fig_BHM_schematic}
\end{figure}

We first try to gain a basic understanding of the BHM to capture the properties of the different phases and transitions between them.
By neglecting correlations between bosons at different sites, the ground state wavefunction within the mean-field (MF) treatment can be described by a product state,
\begin{eqnarray}
\ket{\psi} = \prod_{i} \otimes  \ket{\psi_{i}},
\end{eqnarray}
where $\ket{\psi_{i}}$ describes the state at $i^{th}$ site of the lattice.
For small hopping amplitude $t \ll U$, same integer number of bosons are localized at each lattice site forming a MI phase. In this case, $\ket{\psi_{i}}$ represents the number states $\ket{n}$, describing the MI phase with filling $n$, for which $\langle \hat{a}_{i} \rangle =0$. On the other hand, for larger values of $t$, the bosons delocalize and hop between the different lattice sites, resulting in the formation of an SF phase. In the deep SF regime, $\ket{\psi_{i}}$ can be approximated by a bosonic coherent state $\ket{\alpha}=e^{-|\alpha|^2/2}\sum^{\infty}_{n=0}\alpha^{n}/\sqrt{n!}\ket{n}$, for which the bosonic operators can be represented by classical fields, $\langle \hat{a}_{i} \rangle = \alpha$. 
It is evident that $\langle \hat{a}_{i} \rangle \neq 0$ can distinguish the SF from the MI phase, and therefore serves as an SF order parameter.
Note that, the BHM in Eq.~\eqref{BHM} remains invariant for $\hat{a}_{i} \rightarrow \hat{a}_{i}e^{\imath \theta}$, which is associated with the  U(1) symmetry. The SF phase acquires a definite phase as $\langle \hat{a}_{i} \rangle \neq 0$ due to the spontaneous breaking of this symmetry. 
In general, the two extreme limits can be interpolated by the Gutzwiller wavefunction \cite{Krauth1992},
\begin{eqnarray}
\ket{\psi} = \prod_{i} \otimes \left(\sum_{n} f^{(i)}_{n}\ket{n}_{i}\right),
\end{eqnarray}
where $\ket{\psi_{i}}$ is a linear combination of the number states $\ket{n}_{i}$, $f^{(i)}_{n}$ are Gutzwiller amplitudes determined from the minimization of $\langle \psi | \hat{\mathcal{H}}-\mu\hat{N} | \psi \rangle$, subjected to the constraint $\sum_{n}|f^{(i)}_{n}|^2 = 1$, and $\hat{N}=\sum_{i}\hat{n}_{i}$ denotes the total particle number operator.
Owing to the translational invariance of the homogeneous systems, $f^{(i)}_{n}$ can be considered the same for all lattice sites. For the MI phase with filling $n_{0}$, $f_{n}=\delta_{n,n_{0}}$, whereas for the SF phase, $f_{n}$ are denoted by a linear combination of the different number states, which yields $\langle \hat{a}_{i} \rangle = \sum_{n}f^{*}_{n}f_{n+1}\sqrt{n+1}$.  
Within the Gutzwiller approach, a continuous transition from SF to MI phase can be captured as $t$ decreases \cite{Zoller1998,Sheshadri1993}, which is consistent with the experimental observation \cite{Greiner2002}. 
This transition has been extensively studied using more sophisticated theoretical methods incorporating the quantum fluctuations and correlations \cite{Monien1996, Dupuis2005, Trivedi2009, Dupuis2011, Trivedi1991, Svistunov2007, Svistunov2008, Fabrizio2007, Polak2007}.  
Even a more precise critical behavior of the 2D BHM has been summarized in Ref.\cite{QMC_BHM_precise}, which has also been compared with the existing results.

In a more general prescription, the SF phase in three dimensions is characterized by the ODLRO, as the single-particle density matrix (SPDM) $\rho_{ij}=\langle \hat{a}^\dagger_{i}\hat{a}_{j} \rangle$ approaches a finite value in the limit $|i-j| \rightarrow \infty$ \cite{Yang1962}. This is similar to the non-vanishing SF order parameter $\langle \hat{a}_{i} \rangle \neq 0$, as discussed earlier in the context of MF description. 
However, in lower dimensions, SFs can have a quasi long-range order \cite{Schwartz1977}. 
The Fourier transform of the SPDM yields the momentum distribution \cite{Lewenstein_book2012},
\begin{eqnarray}
n(k) = \frac{1}{\mathcal{N}}\sum_{i,j=0}e^{\dot{\imath}\vec{k}\cdot(\vec{r}_{i}-\vec{r}_j)}\rho_{ij}, 
\end{eqnarray}
where  $\vec{r}_{i}$ represents the lattice vector corresponding to the $i^{th}$ site, $\vec{k}$ is the reciprocal lattice vector (momentum), and $\mathcal{N}$ denotes the number of lattice sites.
The appearance of a strong peak in the momentum distribution $n(k)$ at $k=0$ serves as another signature of superfluidity.
This is a consequence of BEC associated with superfluidity, reflecting a macroscopic occupancy at the $k=0$ state for a homogeneous system. 
When the SF phase of ultracold atoms is released from the optical lattice, other coherent Bragg peaks apart from the prominent central peak at $k=0$ are also visible due to the periodic structure of the lattice, which marks the identification of this phase in experiments \cite{Greiner2002}. On the other hand, such coherent peaks disappear for the MI phase and instead, broadening of the momentum distribution occurs at $k=0$, reflecting the localized nature of this phase.
As pointed out in Ref.~\cite{London,Tisza1938,Tisza1947}, superfluidity is always associated with the condensation of the bosons in the lowest energy state, for which the corresponding condensate fraction \cite{Leggett2001} is defined by $\rho_{c}=\lambda/N$, where $\lambda$ is the largest eigenvalue of the SPDM and $N$ denotes the number of bosons.
Nevertheless, the presence of interactions and correlations can lead to depletion in the condensate, reducing the condensate fraction.
In this context, the condensate fraction in SF $^4$He has been experimentally observed to be very small ($\sim 7-14 \%$) at low temperatures \cite{Glyde2000, Mook1983, Svensson1982, Sokol1989}, which is in stark contrast to the nearly pure condensates of dilute atomic vapors.

From the hydrodynamics point of view, the average density $n_{0}$ of an SF system can be described as a sum of SF and normal components, $n_{0} = n_{S} + n_{N}$ \cite{Landau_superfluid_theory, Stringari_book}.   
The transport property related to the dissipationless flow of the SFs is characterized by the SF fraction $\rho_{s}=n_{s}/n_{0}$ \cite{Leggett1970,Fisher1973_sf}.
This can be quantified from the response of the SF to small perturbations in the phase. 
By imposing a small twist $\Theta \ll \pi$ in the boundary conditions along a particular direction, the SF fraction $\rho_{s}$ can be obtained \cite{Fisher1973_sf,Roth2003},
\begin{eqnarray}
\rho_{s} = \frac{2m}{\hbar^2}\frac{L^2}{N}\frac{(E_{\Theta}-E_{0})}{\Theta^2}, \label{SFF_primary} 
\end{eqnarray}
where $E_{\Theta}$ ($E_{0}$) represent the ground state energy in the presence (absence) of the phase twist, $L$ denotes the system size along which the phase twist is being applied, $N$ is the total number of bosons, and $m$ denotes their mass. 
Alternatively, the above quantity can also be obtained as a response to slow rotation, where the superfluidity is manifested through the Non-classical rotational Inertia (NCRI) \cite{Leggett1970}. This approach has been extensively used in the context of $^4$He \cite{Balibar_review2008}. 

By including long-range interactions between the bosons at different sites with strength $V_{ij}$, the BHM in Eq.~\eqref{BHM} can be extended to the following Hamiltonian,
\begin{equation}
\hat{\mathcal{H}}_{\text{EBHM}} = \hat{\mathcal{H}}_{\text{BHM}} + \frac{1}{2}\sum_{i\neq j}V_{ij}\hat{n}_{i}\hat{n}_{j}. \label{EBHM}
\end{equation}
This is commonly referred to as the extended Bose-Hubbard model (EBHM). In the presence of translational invariance, the long-range interaction can be typically written as $V_{i,j} = V(|i-j|)$, which falls off with increasing distance between the sites $|i-j|$. However, in the case of dipolar interactions, the long-range interaction can be anisotropic in nature depending on the orientation of the dipoles, which can give rise to a rich variety of density-ordered phases \cite{Zoller2010, Kawashima2012, Rey2015, Morigi2024, Santos2002, Sun2007, Danshita2009, Sansone2014, Zhang2021, Zhang2022, Macia2012, Bombin2017}. 
In this context, the Rydberg atoms are promising candidates to study such density-ordered phases due to the presence of long-range interactions.
Moreover, all-to-all interactions can also be engineered in cavity setups \cite{Cavity_QED_review} [see for example the schematic in Fig.~\ref{Fig_BHM_schematic}].

\begin{figure}
	\centering
	\includegraphics[width=\columnwidth]{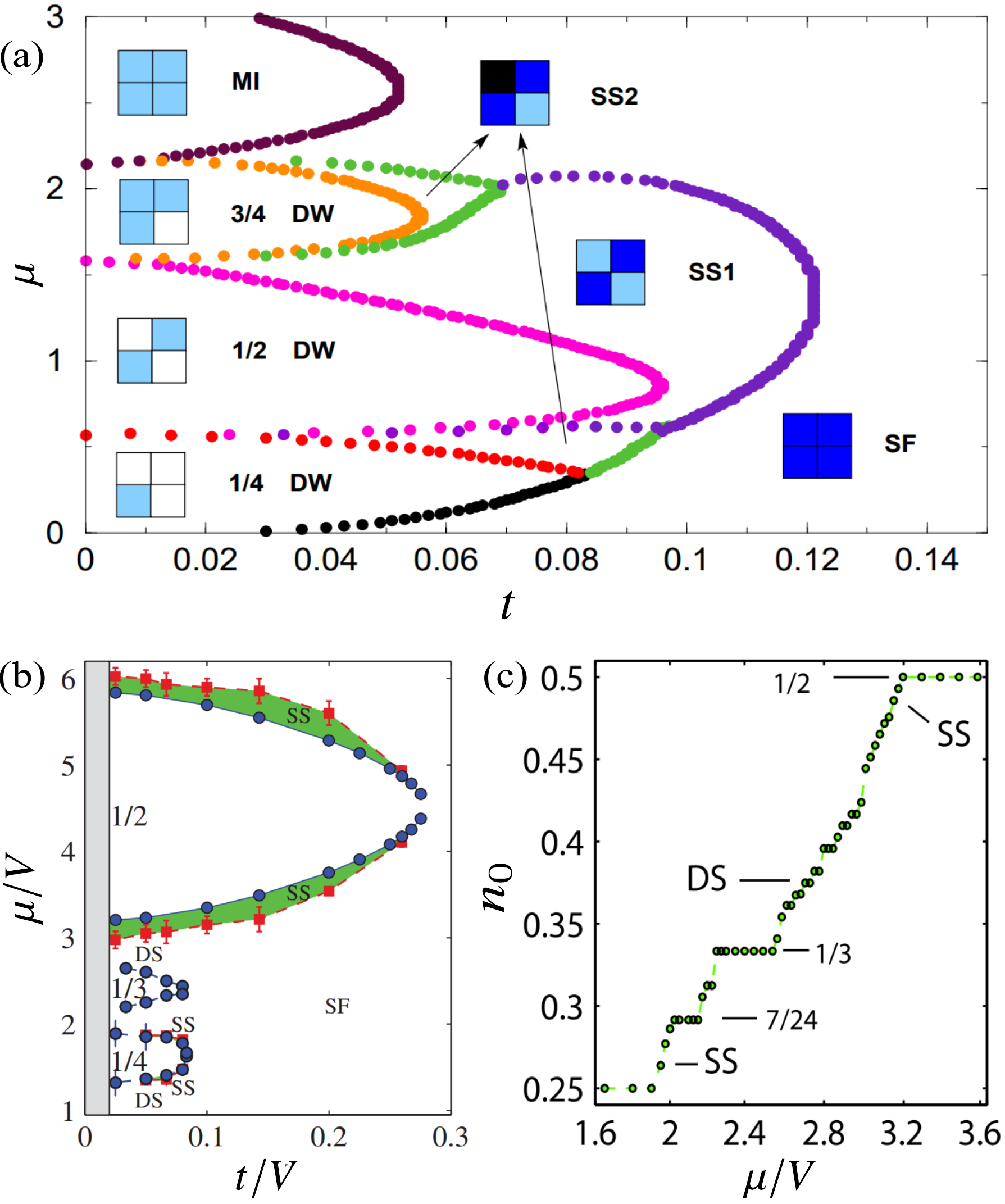}
	\caption{Zero temperature phase diagram on a square lattice.
	(a) Phase diagram of softcore bosons obtained using Gutzwiller mean-field (MF) theory. The insets denote the schematics of density wave (DW), Mott Insulator (MI), supersolid (SS), and superfluid (SF) phases on a unit cell. (b) Phase diagram of hardcore bosons in the presence of dipolar interactions with strength $V$ obtained using Quantum Monte Carlo (QMC) simulations. The different lobes denote the following: solids (DW) with various fillings, SF, SS, and {\it Devil's staircase}. (c) Solids and SS for a system with $t/V = 0.05$. Figure (a) is reproduced from Ref.~\cite{Sinha2005}. $\copyright$ IOP Publishing Ltd. All rights reserved. Figures (b) and (c) are reprinted figure with permission from Ref.~\cite{Zoller2010}, Copyright (2010) by the American Physical Society.}      
\label{phase_diagram_supersolids}
\end{figure}

\begin{figure*}
	\centering
	\includegraphics[width=0.9\textwidth]{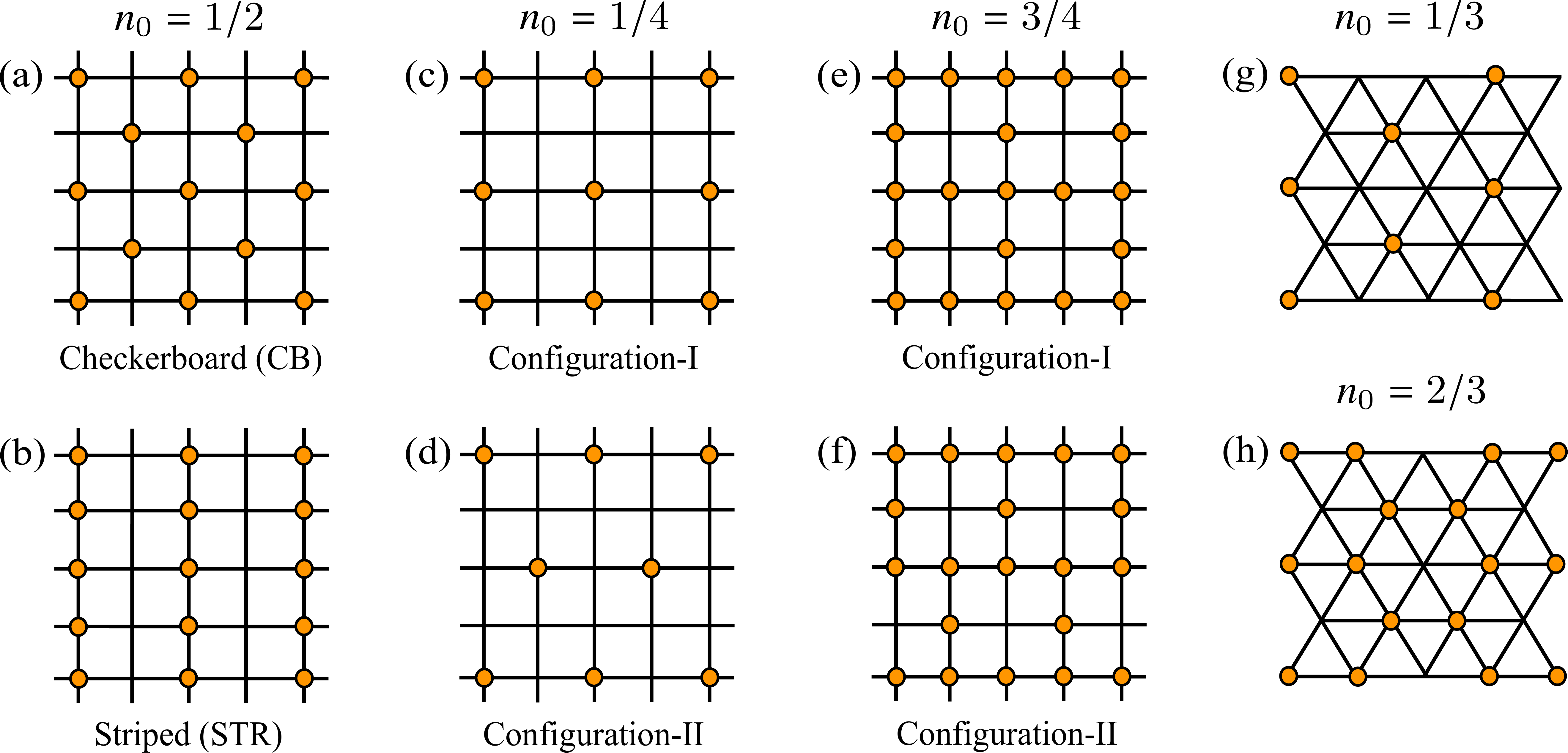}
	\caption{Schematics of different solids or density wave (DW) phases of hardcore bosons in EBHM with nearest (NN) and next-nearest neighbor (NNN) interactions, $V_{1}$ and $V_{2}$, respectively, charted out for different fillings $n_{0}$ in the (a-f) square lattice and in the (g,h) triangular lattice. 
}      
\label{Fig_solid_schematic_hc}
\end{figure*}

To understand the mechanism behind the formation of SSs, we focus on a minimalistic EBHM model with only the nearest-neighbor (NN) interaction term $V_{1}\hat{n}_{i}\hat{n}_{i+\delta}$ in Eq.~\eqref{EBHM}, where $\delta$ represents the NNs of the $i^{th}$ site. In the atomic limit ($t=0$), at half-filling ($n_{0}=1/2$), the `checkerboard' (CB) like density-ordering is energetically favored in a square lattice.
Within the MF treatment, such density-ordered phase can be described by a two sub-lattice structure on the unit cell as a consequence of the breaking of discrete translational invariance. 
Moreover, the reflection of periodic modulation in the SPDM $\rho_{ii} = \langle \hat{a}^\dagger_{i} \hat{a}_{i} \rangle$, indicates a DLRO, which can be quantified from the {\it static structure factor} \cite{Lewenstein_book2012},
\begin{eqnarray}
S(k) = \frac{1}{N}\sum_{i,j}e^{\dot{\iota}\vec{k}\cdot(\vec{r}_i-\vec{r}_j)}\langle \hat{n}_{i}\hat{n}_{j}\rangle,
\end{eqnarray}
representing the Fourier transform of the density-density correlations between bosons at different sites. 
The above quantity displays pronounced peaks at the reciprocal lattice vectors as the intensity of the light scattering off the crystal is maximum corresponding to these points, which can be used for characterizing the different density orderings.
With increasing the hopping strength $t$, it is expected that the SF order will emerge as the particles are delocalized. For hardcore bosons ($n_{i} \in \{0,1\}$ as $U \rightarrow \infty$), this model can be mapped to a `quantum lattice gas' model, which was originally used for studying the solid to SF transition in $^4$He \cite{Matsuda_QLG1956,Matsuda_QLG1957}. 
It is important to mention that, although the presence of only the NN interaction yields a density-ordered phase with CB ordering, the SS phase still remains absent for hardcore bosons.
Interestingly, by incorporating interactions beyond the NNs, it is possible to realize an SS phase in between the solid and SF phases \cite{Fisher1973}.
In the presence of sufficiently strong NNN interactions, a stripe (STR) ordering is also favored in addition to the CB ordering on a square lattice. While both the CB and STR ordered SS phases of hardcore bosons are obtained within the MF approach, Quantum Monte Carlo (QMC) studies have shown that the CB SS is thermodynamically unstable and undergoes a phase separation to density wave (DW) and SF states \cite{Batrouni2000}; however, the STR SS can still remain as a stable phase \cite{Troyer2001}.
In contrast, the QMC simulations for hardcore bosons with realistic dipolar interactions indicate that it is possible to realize a stable SS phase with CB ordering in between the solid and the SF states \cite{Zoller2010}, as depicted in the phase diagram in Fig.~\ref{phase_diagram_supersolids}(b).
Furthermore, apart from the inclusion of longer range interactions, the presence of an NNN anisotropic hopping can also lead to supersolidity in hardcore bosons \cite{anisotropic_hopping}.
The models for hardcore bosons are also equivalent to spin-1/2 systems \cite{Matsuda1970}, where the supersolidity corresponds to the coexistence of a staggered and anti-ferromagnetic ordering \cite{Fisher1973}.
The search for supersolid phases in different types of spin systems (as well as hardcore bosons) has also been explored using various techniques \cite{Zoller2010, Kawashima2012, Rey2015, Morigi2024, Batrouni2000, Troyer2001, anisotropic_hopping, Otterlo1995, Batrouni1995, Scalettar1995, Frey1997, Frey1998, Sengupta2007, Melko2008, Pollet2008, Kwon2010, Kwon2010_JPCM, Danshita2012, Pupillo2019, Pupillo2020, Wu2020, Prestipino2021, MalakarJSM2023}. 
For bosons with large filling (with occupancy larger than unity), the EBHM maps onto a quantum phase model (QPM), which has been studied in the context of arrays of coupled Josephson junctions \cite{JJ_array_review2001}. These systems can also host such exotic phases \cite{Roddick1993, QPM1994, Roddick1995, Fazio1997, Frey1997}. 

On the other hand, realistic systems with both the on-site and long-range interactions hold promise for observing the SS phase \cite{Santos2002, Sun2007, Danshita2009, Sansone2014, Zhang2021, Zhang2022}, as the NN interaction is sufficient for the emergence of a coexisting order \cite{Sinha2005, Pinaki2005, Demler2006, Iskin2011, Kimura2011, Kawashima2012_2, Suthar2019, Suthar2020}. 
For softcore bosons on a square lattice, the presence of both the NN and NNN interaction reveals the existence of various density-ordered and supersolid phases within the mean-field approach \cite{MalakarJSM2023,Sinha2005} [see the phase diagram in Fig.~\ref{phase_diagram_supersolids}(a)].
Furthermore, the QMC studies have confirmed that, even with only the NN interactions, a stable SS phase indeed appears for softcore bosons \cite{Pinaki2005}. 
However, in the hole doped region, such an SS phase can become unstable and undergo a phase separation into a solid and an SF. 
For a homogeneous system, the cluster SS phase has been obtained using the QMC technique \cite{Cinti2010, Saccani2011, Moroni2012, Boninsegni2012, Pohl2013, Rossi2013, Lechner2015, Kora2019, Ciardi2024, Ciardi2025}. Within the classical field approximation, dilute homogeneous gases even in the presence of long-range interactions can be described by the GPE [see also the appendix~\ref{appendix1}]. 
Moreover, the excitations can also be obtained from the linear fluctuation analysis, which is equivalent to the Bogoliubov quasi-particle energies \cite{Stringari_book}. However, in the experimental observation of supersolidity in dipolar condensates, in addition to the long-range interactions, the beyond mean-field effects, known as the LHY correction \cite{LHY_original_paper1,LHY_original_paper2}, also play a key role in the formation of the SS phase.
These phases and their excitations have been theoretically studied by extending the GPE with the LHY correction \cite{Blakie2018, Pohl2019, Pohl2021, Bissett2021, Langen2022, Adhikari2022, Adhikari2022_2, Rejish2022, Adhikari2023, Ripley2023, Blakie2023, Ferlaino_expt_gapless_mode_SS2019, Stringari_sound_2024, Blakie_exc_spec_2024, Blakie_exc_spec2_2024, Blakie2025}. Alternatively, the three-body interaction can also lead to stable SSs in bilayer systems \cite{Shlyapnikov2015}.

In the next section, we discuss the SS phases characterized by the different kinds of density orderings at zero as well as finite temperatures.

\section{Density ordering in supersolids and finite temperature effects}
\label{density_orderings_in_supersolids}
In this section, we provide an overview of the various possible density-orderings that arise in lattice-SSs due to long-range interactions. Even when the hopping amplitude is small, the long-range interaction between the particles, their density, as well as the underlying lattice structure, play a crucial role in the formation of solids with different fillings. The SSs that stem from the corresponding solids can retain their respective density-orderings. Furthermore, they can also undergo a structural transition by tuning the interactions and the temperature.
On the other hand, for continuous systems, a triangular structure of the density modulation is energetically favored, which has been shown for the SS phase of a dipole blockaded gas using QMC technique \cite{Cinti2010}. From the analysis of the GPE around the wave-vector corresponding to the `roton' instability, the hexagonal and bcc (or hcp) SS structures have also been obtained in 2D and 3D, respectively \cite{Pomeau_Rica1994}.
However, for realistic systems, the trap geometry as well as the anisotropic nature of the dipolar interaction also play an important role in determining the structure of the density modulation \cite{Blakie2018, Pohl2019, Pohl2021, Bissett2021, Langen2022, Adhikari2022, Adhikari2022_2, Adhikari2023, Rejish2022, Ripley2023}. For example, in the presence of a harmonic confinement along the polarization of dipoles, the triangular structure of dipolar SSs can undergo structural transitions to exhibit stripe as well as honeycomb patterns by tuning the short-range and dipolar interactions \cite{Pohl2019, Ripley2023}.

For the lattice systems, in the absence of hopping, the long-range interaction results in the formation of solids with increasing filling, exhibiting a {\it devil's staircase} like structure \cite{Zoller2010, Morigi2024}. 
To understand the formation of the ordered phases, we consider the case of a 2D square lattice as an example, where the long-range interaction is truncated till the NNN term in Eq.~\eqref{EBHM}, with the strength of NN and NNN interaction being $V_{1}$ and $V_{2}$, respectively. 
At half-filling ($n_{0}=1/2$), there can be two different kinds of ordering depending on the relative strength $V_{2}/V_{1}$, namely, the CB ($2V_{2}/V_{1}<1$) and STR ($2V_{2}/V_{1}>1$) ordering. However, for hardcore bosons, unlike the CB solids, the STR solids give rise to a stable SS phase \cite{Troyer2001}. 
Such ordering can be detected from the static structure factor $S(\vec{k})$, which exhibits peaks at $\vec{k}=(\pi,\pi)$ for the CB, and at $\vec{k}=(\pi,0)$ or $(0,\pi)$ for the STR ordering.
Interestingly, for $2V_{2}>V_{1}$, in addition to the STR solid phase, other solids with filling $n_{0}=1/4$ and $n_{0}=3/4$ can also form in the hole and particle dominated regions, respectively. The different possible DW phases in the square and triangular lattice are illustrated in Fig.~\ref{Fig_solid_schematic_hc}. 
While there can be different configurations that are degenerate or closely spaced in energy, the inclusion of either true long-range interactions or quantum fluctuations can favor a particular ordering.
For instance, the solid phases with fillings $n_{0}=1/4$ and $n_{0}=3/4$ each have two degenerate configurations, as shown in Fig.~\ref{Fig_solid_schematic_hc}(c-f). 
However, the inclusion of longer range interactions lowers the energy of configuration-II \cite{Danshita2012}, see Fig.~\ref{Fig_solid_schematic_hc}(d,f).
Even in the presence of only the NN repulsion, depending on the geometry of the lattices, there can be different density-ordered phases, such as $n_{0}=1/3$ and $n_{0}=2/3$ filled solids in a triangular lattice, depicted in Fig.~\ref{Fig_solid_schematic_hc}(g,h). In this case, the frustration around half filling favors the formation of a SS phase \cite{Murthy1997, Troyer2005, Damle2005, Melko2005, Prokofev_triangular2005, Hassan2007, Moessner2008, Pollmann2009, Pollet2010, Zhang2011, Danshita_triangular2012, Pupillo2016, Pelster2016, Wen2007, Malakar_triangular2020}, which is discussed later in Sec.~\ref{frustration}.

For softcore bosons, there can be more complications in the density-ordering due to the interplay between the repulsive on-site interaction strength $U$, as well as NN and NNN interactions, giving rise to a diverse range of solid phases. 
When the on-site interaction strength $U$ is sufficiently large compared to $V_{1}$ and $V_{2}$, the phase diagram resembles to that of hardcore bosons (as discussed in Ref.~\cite{MalakarJSM2023}), where mainly the CB and STR density ordering dominate. On the other hand, when $U$ is relatively weak ($U\lesssim{\rm min}(2V_{1},4V_{2})$), the bosons prefer to stack on a particular site of the unit cell. This regime is preferable for the formation of more stable SSs, which occupy a larger region in the phase diagram \cite{MalakarJSM2023,Iskin2011}. In the intermediate regime, when $U$ is comparable with both the NN and NNN interaction strengths, a plethora of solid phases can arise with different configurations. For example, when CB ordering is favored ($V_{2}<2V_{1}$), different types of solids are charted in Fig.~\ref{Fig_solid_schematic_sc}.

\begin{figure}
	\centering
	\includegraphics[width=\columnwidth]{Fig_schematic_DW_sc.pdf}
	\caption{Different solid (DW) phases of softcore bosons in the EBHM with NN and NNN interactions in the square lattice. (a) The mean field phase diagram in $\mu/V_{1}$–$U/V_{1}$ plane illustrating the different insulating (DW) phases close to the atomic limit ($t/V1 \sim 0$), in the regime where NN interaction dominates ($2V_{2}<V_{1}$). The vertical dotted lines marked by the solid arrows represent the critical boundaries of the different regimes of interaction strengths. (b) Schematics of the different solids represented by the density configuration $(n_{1},n_{2},n_{3},n_{4})$ of the unit cell. The unit cells of DW-3/4 phases are illustrated as an example in (b). Figure (a) has been reproduced from Ref.~\cite{MalakarJSM2023}. $\copyright$ IOP Publishing Ltd. All rights reserved.}      
\label{Fig_solid_schematic_sc}
\end{figure}

Identifying the density modulated structures in the presence of true long-range interactions is relevant for understanding the formation of stable SSs and their characterization.
The existence of various SS phases of dipolar bosons in lattices has been predicted using different techniques \cite{Zoller2010, Kawashima2012, Rey2015, Morigi2024, Santos2002, Sun2007, Danshita2009, Sansone2014, Zhang2021, Zhang2022}.
For instance, the phase diagram obtained using the classical field method exhibits stable SS phases with both the CB and STR ordering in a square lattice, which can undergo a structural transition by changing the orientation of the dipoles \cite{Danshita2009}. 
As discussed earlier, even truncating the long-range interaction up to the NNN term gives rise to a diverse range of solids with different configurations that are energetically close. Extending beyond the NNN term, true long-range interactions can lead to the formation of metastable solid states with various filling that are quasi-degenerate, and can even compete with the ground state, similar to a disordered system \cite{Trefzger2007, Trefzger2008, Malomed2012, Rieger2018}. 
Moreover, a recent experiment with dipolar bosons in an optical lattice also reported the emergence of metastable STR solids \cite{Griener2023}. Therefore, it is also interesting to explore the formation of SS phases as metastable states \cite{Modugno_expt_SS2019, Zhu2023, Zhang2024, Blakie2025}. 

\begin{figure}
	\centering
	\includegraphics[width=0.9\columnwidth]{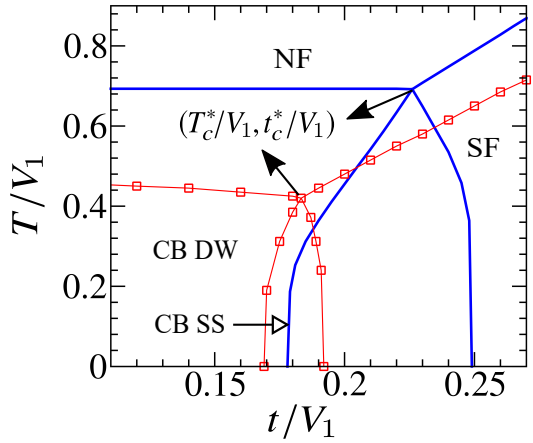}
	\caption{Finite temperature phases of softcore bosons in the $T/V_{1}$-$t/V_{1}$ plane within mean-field (blue lines) and cluster mean-field (red lines with squares) theories for fixed values of chemical potential $\mu$ around the half filled region. The tetra-critical point $(T^{*}_{c}/V_{1},t^{*}_{c}/V_{1})$ is surrounded by the four phases: normal fluid (NF), superfluid (SF), checkerboard supersolid (CB SS), and checkerboard solid (CB DW). The boundaries denote continuous transitions between the different phases. Figure is reproduced from Ref.~\cite{MalakarJSM2023}. $\copyright$ IOP Publishing Ltd. All rights reserved.}      
\label{Fig_finite_temp_sc}
\end{figure}
 
In this context, the effect of thermal fluctuations can also play a key role in determining the stability of the SS phases. 
The possibility of finding a stable SS phase at finite temperature was first considered within the lattice gas model \cite{Fisher1973}. Later, the effect of thermal fluctuations has been studied within the MF, cluster mean-field (CMF), and QMC simulations \cite{MalakarJSM2023, Malakar2023, Kawashima2012, Rey2015, Troyer2002, Troyer2004, Laflorencie2007, Kwon2010_thermal, Kawashima2012_2, Miyashita2009, Suthar2022, Abouie2020, Bombin_melting2019, Prokofev_triangular2005, Hassan2007, Pollet2010, Pupillo2016, Pelster2016, Malakar_triangular2020}. 
Similar to the thermal transition of solid and SF phases to a normal fluid (NF) \cite{Trivedi2011, Micnas2015, Ray2022}, the SSs also melt with increasing temperature.
However, there can be different melting pathways for SSs, where either the density or SF order can disappear first, forming an SF or DW state, respectively \cite{Laflorencie2007, Kwon2010_thermal, MalakarJSM2023, Miyashita2009}. The finite temperature phase diagrams obtained within the MF and CMF approach [shown in Fig.~\ref{Fig_finite_temp_sc}] reveal the existence of a tetra-critical point surrounded by the different phases. The effect of correlations suppress the melting temperature of the SSs, making the observation of such correlated phases a delicate issue. 
Apart from determining the melting pathways of the SS phases, the thermal fluctuations can also induce structural transitions between them \cite{MalakarJSM2023}.
Such thermal effects are relevant to the formation of the SS phase, as recent experiments have demonstrated the emergence of supersolidity in dipolar condensates through the technique of direct evaporation \cite{Ferlaino_expt_SS2019, Ferlaino_expt_SS2021, Ferlaino_expt_SS2022}.
Furthermore, it has been observed in an experiment that the melting of dipolar fluids can result in the formation of SSs where the density ordering emerges with increasing temperature  \cite{Ferlaino_nat_comm2023}. This phenomenon can crucially depend on the nature of the SS-SF phase boundary at finite temperatures.

In the next section, we chart out the different directions for achieving the SS phases.

\section{Different directions for achieving supersolidity}
\label{Different_directions_for_achieving_supersolidity}
For a long time, $^4$He was considered to be the ideal candidate for observing the SS phase \cite{Prokofev_review2007, Balibar_review2008, Balibar_review2010, Prokofev_review2012, Reatto_review2013, Yukalov_review2020}. However, advancements in quantum technologies have shifted the focus to other platforms, such as ultracold atomic setups, which have proven to be promising alternatives for exploring such phases \cite{Bottcher_review2021, Stringari_review2023}. In this section, we outline the different mechanisms through which the SS phases can appear in systems beyond $^4$He.

\subsection{Bosons with long-range interactions}
\label{bosons_with_lr_interactions}
As discussed earlier in Sec.~\ref{density_orderings_in_supersolids}, the long-range interactions serve as a key ingredient for different density orderings, which are an essential component for the formation of the SS phases \cite{Ferlaino_review2023}. In this context, the dipolar atoms and polar molecules are considered to be the most suitable candidates for observing the SS phase, where the orientation of the dipoles can play an important role in determining its stability. Recent experiments with dipolar condensates of $^{166}$Er and $^{164}$Dy atoms \cite{Pfau_expt_SS2019, Modugno_expt_SS2019, Ferlaino_expt_SS2019, Ferlaino_expt_SS2021, Ferlaino_exptnature_SS2021, Ferlaino_expt_SS2022} reveal the signature of such fascinating state of matter, although its formation is a much more delicate issue where the quantum fluctuations play a significant role, preventing the collapse of the condensate by forming quantum droplets. The phase coherence between the droplets builds up the supersolidity of the {\it coherent droplet} state \cite{Giamarchi_expt_SS2021}. Furthermore, loading dipolar atoms in an optical lattice can avoid the collapse and lead to the formation of strongly correlated SS phases, which have been explored theoretically \cite{Zoller2010, Kawashima2012, Rey2015, Morigi2024, Santos2002, Sun2007, Danshita2009, Sansone2014, Zhang2021, Zhang2022}. A recent experiment on the Bose-Hubbard model with $^{168}$Er atoms already exhibits the emergence of different solids \cite{Griener2023}.

Alternatively, the ultracold Rydberg atoms can also generate a long-range interaction, $V(r_{ij}) = C/|\vec{r}_i-\vec{r}_j|^6$, which can be utilized for density ordering. 
The theoretical studies indicate the possibility of finding the SS phase of Rydberg dressed Bose-Einstein condensates (BEC) \cite{Cinti2010, Pohl_Rydberg2010, Pohl_Rydberg2012, chiral_SS_SOC1, Lyu2020, Wang2022, chiral_SS_SOC2} as well as Rydberg atoms in an optical lattice \cite{Eggert2013, Kush2014, Hofstetter2017, Hofstetter2018, Hofstetter2018_quasiparticle_spectraSS, Hofstetter2019_non_eqSS, Hofstetter2019, Hofstetter2022, Pollet_Rydberg2025}. Although the Rydberg excitations have a finite lifetime, the SS phase can be observed as a non-equilibrium state of such open quantum systems \cite{Hofstetter2019_non_eqSS}. Moreover, the appearance of density ordering has already been reported in experiments with Rydberg dressed spin lattices \cite{Pohl_Rydberg2016}.

\subsection{Atoms in optical cavities}
\label{atoms_in_optical_cavities}
Ultracold atoms coupled to optical cavities have become a new platform to probe the exotic phases of quantum matter \cite{Cavity_QED_review}, particularly due to the cavity-mediated effects. 
A SS phase emerges when a BEC cloud is symmetrically coupled to modes of two optical (crossed linear) cavities, as the continuous translational symmetry breaks along a particular direction, which is monitored through cavity photons, while the phase coherence ensures superfluidity \cite{Esslinger_expt_2017, Esslinger_expt2_2017}.
Similar SS phases of BEC have also been realized in a ring cavity \cite{Zimmerman_expt_2020, Ritsch2018, Ritsch2023, Deng2023}.

Furthermore, BEC coupled to an optical cavity driven by a pump laser reveals a lattice-SS phase, where self-organized density modulations appear, accompanied by a superradiant transition \cite{Esslinger_expt_2010, Hemmerich_expt_2015}.
Superimposing an optical lattice in this setup [see for example Fig.~\ref{Fig_BHM_schematic}] opens up another possibility to investigate the effect of correlations in these many-body systems, where apart from the usual SF and MI phases, a lattice-SS phase has also been theoretically studied \cite{Eggert2013, Rieger2018, Hofstetter2013, Mehkov2016, Dogra2016, Rieger2016, Batrouni2017, Panas2017, Zheng2018, Kraus2019, Kraus2022, Guan2019, Carl2023, Ray2024} and observed in experiment \cite{Esslinger_expt_2016} due to the competition between the short- and cavity-mediated long-range interactions.
In these setups, such cavity-mediated infinite-range interaction plays a crucial role in the formation of the SS phase with density modulations in the condensates at both zero and finite temperatures, which has been investigated in Refs.\cite{infinite_range_interaction1, infinite_range_interaction2, infinite_range_interaction3}.
Furthermore, dipolar atoms loaded in a 2D optical lattice inside a cavity can enhance the stability of the SS phase, due to the combined effect of dipolar and cavity-mediated long-range interactions \cite{cavity_dipolar_bosons1,cavity_dipolar_bosons2}.
Natural dissipative processes in these systems, such as photon loss, make them an ideal platform for exploring fascinating out-of-equilibrium phases \cite{Cavity_QED_review, Esslinger_dissipative_phasesPRX}.

Due to the advancement of the cavity and circuit QED setups, it has become possible to realize an array of coupled cavities containing atoms \cite{Cavity_dimer, Cavity_array}, which is a promising candidate to realize the crystalline and SS phases of photons \cite{Fazio_solidphoton2013, Fazio_solidphoton2015, Nori2015, Cole2014, Guo2019, Scott2021}.

\subsection{Supersolidity using photons}
\label{supersolid_photons}
Apart from ultracold atoms, supersolidity has also been achieved using photons in a recent remarkable experiment \cite{nature_SS_photons, PRL_SS_photons}. In this work, a fascinating SS phase was realized through a driven dissipative exciton-polariton condensate in a photonic crystal waveguide. The density modulations in the polaritonic state, indicating the breaking of translational symmetry--a hallmark of supersolidity--have been measured with high precision. Furthermore, the local coherence of the wavefunction ensures the simultaneous existence of superfluidity. This approach offers a promising route to investigate SS phases in driven dissipative non-equilibrium systems.

\subsection{Spin-orbit coupling}
\label{SOC_supersolid}
With the advancement of quantum engineering techniques, it has become possible to experimentally generate synthetic spin-orbit coupling using the internal states of the cold atoms \cite{SOC_BEC2011, Spielman_SOC_review, Stringari_SOC_BEC_review2015, SOC_BEC_review2015, SOC_BEC_review2016}. As a result of the modified dispersion, such spin-orbit coupled condensates can form a STR phase due to the superposition of the two momentum states within certain parameter regimes \cite{SOC_BEC2011, Stringari_SOC2012}. This phase exhibits a periodic density modulation, which can be controlled by the strength of the spin-orbit coupling \cite{Spielman2011, Ketterle2017, Leticia_expt_2024}. Since such super-stripe phase is a BEC with coexisting density modulation, it can be considered as a SS phase, although its mechanism is different from the original ideas of supersolidity. 
Theoretical studies also reveal the possibility of more exotic phases such as chiral, spiral, and striped SSs \cite{chiral_SS_SOC1, Lyu2020, Wang2022, chiral_SS_SOC2, Zhu2023, Liu2020, Peng2020, Bai2024, Ritsch2021_SOC, Zhu2024, Ritu2024}.

\subsection{Frustration}
\label{frustration}
Frustration is a well-known phenomenon that occurs in spin-1/2 systems (hardcore bosons), particularly due to the underlying structure of the lattice, which can lead to formation of various novel phases \cite{Mila2011, Balents2010, Gingras2001}. 
While the SS phase is absent for hardcore bosons in a square lattice with only NN interactions, it can appear in a frustrated triangular lattice \cite{Murthy1997, Troyer2005, Damle2005, Melko2005, Prokofev_triangular2005, Hassan2007, Moessner2008, Pollmann2009, Pollet2010, Zhang2011, Danshita_triangular2012, Pupillo2016, Pelster2016, Malakar_triangular2020}.
A simple classical analysis of the spin system indicates the possibility of such an SS phase \cite{Murthy1997}, which was later confirmed by using more sophisticated techniques \cite{Troyer2005, Damle2005, Melko2005, Prokofev_triangular2005, Hassan2007, Moessner2008, Pollmann2009, Pollet2010, Danshita_triangular2012, Pupillo2016, Pelster2016, Malakar_triangular2020}. Moreover, for softcore bosons, the competition between the on-site interaction and frustration can also lead to the emergence of interesting phases, especially the stacked SSs \cite{Wen2007, Malakar_triangular2020}. In addition, the effect of the interplay between frustration and thermal fluctuations on the stability of the SSs has also been studied \cite{Prokofev_triangular2005, Hassan2007, Pollet2010, Pupillo2016, Pelster2016, Malakar_triangular2020}. 

\subsection{Pairing}
\label{pairing} 
The ultracold atomic setups are ideal platforms to search for even more exotic states of matter such as paired phases of bosons \cite{Daley2006_pairing, Folling2007, Lukin2013}. In particular, a paired SF state can appear due to attractive interaction between single or two-component bosons \cite{Diehl2009, Yang2010, Yang2011, Wessel2011, Dutta2015, Tapan2018, Kuklov2004, Hu2009, Menotti2010, Menotti2022} similar to superconductors; however, their stability may require additional constraints (like restriction on site occupancy) \cite{Diehl2009, Yang2010, Yang2011, Wessel2011, Dutta2015}. 
Apart from these, other mechanisms like pair hopping can also be engineered to induce pairing between the bosons \cite{Ripoll2008, Zhang2009, Hauke2012, Sandvik2019}, which can be described by the effective Hamiltonian,
\begin{eqnarray}
\hat{\mathcal{H}}_{p} = -t_{p}\sum^{{\rm NN}}_{\langle i,j \rangle}\left(\hat{a}^{2\dagger}_{i}\hat{a}^2_{j} + {\rm h.c}\right), \label{pair_hopping_ham}
\end{eqnarray}
where $t_{p}$ denotes the effective pair hopping amplitude.
Importantly, the paired phases are characterized by the order parameter $\langle \hat{a}^2 \rangle \neq 0$, while $\langle \hat{a} \rangle=0$.
As a natural extension, the inclusion of beyond on-site, longer range interactions in these systems can generate density ordering in the paired SF phase, leading to the formation of a novel paired SS state \cite{Wang2013, Wang2013_2, Malakar2023}.
The incorporation of the pair hopping term [Eq.~\eqref{pair_hopping_ham}] in EBHM [Eq.~\eqref{EBHM}] with NN and NNN interaction can give rise to a paired SS phase along with the usual SSs \cite{Malakar2023}. Moreover, even with just the NN interaction, the combination of pair hopping and frustration in a triangular lattice can also generate paired SSs \cite{Wang2013, Wang2013_2}.
At finite temperatures, such paired SS eventually melts to the NF phase, but as discussed earlier, there can be two melting routes similar to atomic SSs, either through the paired DW or paired SF phase \cite{Wang2013, Malakar2023}.
In this regard, the dipolar bosons in bilayer systems are also promising candidates for achieving such paired phases, where the anisotropic nature of the long-range dipolar interactions can result in interlayer attraction and intralayer repulsion, facilitating the formation of such exotic phases \cite{Santos2007, Astrakharchik2014, Sinha2015, Trefzger2009, Pupillo2013}.
Other than cold atoms, pairing can also be realized in cavity and circuit QED setups \cite{Mora2019, Guo2019, Tapan2020}, making them a suitable platform for observing the different paired phases.
Furthermore, in the case of a superconductor, electrons (fermions) can pair up to form composite particles called Cooper-pairs which develop phase coherence like bosons leading to a dissipationless flow similar to SFs.
In this context, it is also worth mentioning that the Fulde–Ferrell–Larkin–Ovchinnikov (FFLO) state of fermionic SF in the presence of interspecies imbalance can give rise to a periodic structure in the pairing order parameter \cite{FF,LO} similar to the paired SS phase.
Apart from the solid state platforms \cite{solid_state_platform1, solid_state_platform2}, such modulated SF phase can also be achieved in cold fermionic atoms \cite{FFLO_review, FFLO_expt, Orso2007, LO_superfluidity}. 

\subsection{Binary mixtures}
\label{binary_mixtures}
Theoretical studies on Bose-Fermi mixtures reveal that, despite the presence of only short-range interactions, fermions can induce an effective long-range interaction in the bosonic component, leading to the formation of a density-ordered SS phase \cite{Blatter2003, Snoek2008, Roy2008, Hur2009, Sinha2009, Gajda2025}. Such a SS phase can even be observed in two-component bosonic mixtures \cite{Hofstetter_two_comp2009, Ichinose2014, Chen2014, Ralko2014, Wilson2016, Abouie2017, Saito2022, Bland2022, Majumder2023, Bland2023, Majumder2024}.

Although the SS phase can be achieved in different setups, its universal characteristic features can be captured from the collective excitations, which we discuss in the next section.

\section{Signature of supersolids from Collective excitations}
\label{signature_of_supersolids_from_collective_excitations} 
\begin{figure}
	\centering
	\includegraphics[width=\columnwidth]{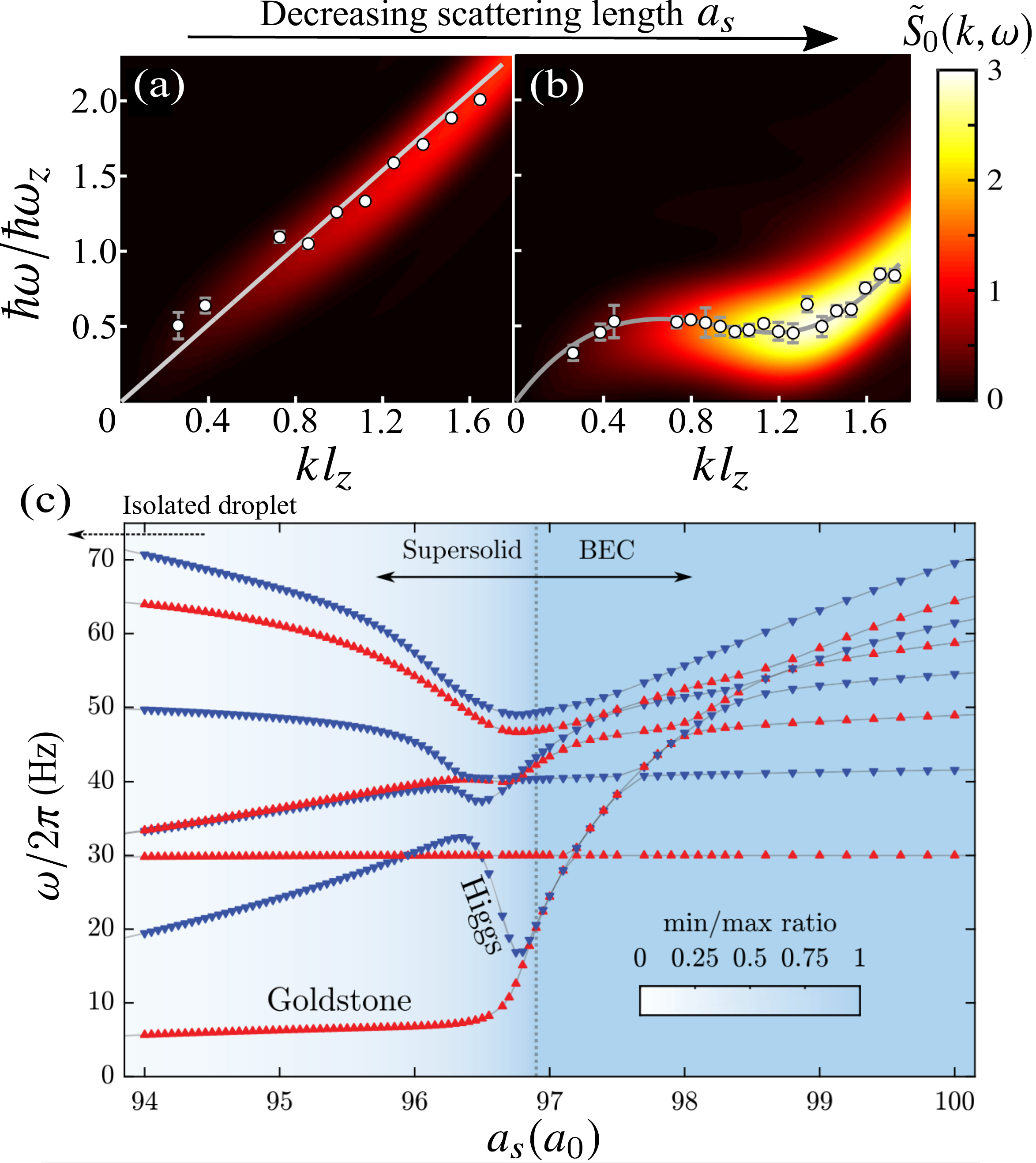}
	\caption{Excitations of dipolar gas: (a,b) Roton mode softening in dipolar BEC of $^{166}$Er atoms. The experimentally extracted excitation spectrum (white dots) is compared with that obtained from Bogoliubov theory (solid line). Colormap denotes the dynamical structure factor $\tilde{S}(k,\omega)$, which is enhanced when the roton mode softening occurs. (c) The excitation frequencies $\omega/2\pi$ of the eight lowest Bogoliubov modes as a function of the scattering length $a_{s}$. Red (blue) triangles indicate that the density variation function of the corresponding mode has odd (even) parity. The background color indicates the ratio between density in between central and side droplets and peak density of the center droplet as a measure of the SF fraction. The vertical dashed line indicates the position of the phase transition. Figures (a) and (b) are reprinted figure with permission from Ref.~\cite{Ferlaino_roton2019}, Copyright (2019) by the American Physical Society. Figure (c) is reprinted figure with permission from Ref.~\cite{Pfau_Higgs_mode}, Copyright (2019) by the American Physical Society.}      
\label{Fig_exc_spec_dipolar_gas}
\end{figure}

The nature of excitations also reveals distinct phases, as they reflect the underlying order and symmetry-breaking. The emergence of a new phase is often signaled by the development of a soft mode, indicating an underlying instability.
In SF $^4$He, the excitations exhibit a unique dispersion with a roton-maxon structure \cite{Landau_spectrum, Feynmann_spectrum, Cowley1970, Donnelly1981}, and it was believed that the roton minimum at finite momentum could act as a precursor to the emergence of a periodically modulated structure, leading to the formation of a SS phase \cite{Nozieres2004}.
Although such a phase has not yet been experimentally observed in $^4$He in a conclusive way, the appearance of a roton minimum in the SF has been found to be associated with the emergence of a density ordering, which has been observed in other platforms like cold atoms and cavity setups \cite{Lyu2020, Ritsch2023, Ferlaino_expt_gapless_mode_SS2019, Macia2012, Sinha2005, MalakarJSM2023, Malakar2023, Lewenstein2003, Kurizki2003, Martone2012, Ferlaino_roton2018, Ferlaino_roton2019, Pfau_roton2021, Pfau2_roton2021, Ripley2023, Stringari_sound_2024, Esslinger_roton2012, Busch2014, Deng2015}. In the case of dipolar condensates, such mode-softening phenomenon has been studied theoretically and has also been observed in recent experiments \cite{Lewenstein2003, Kurizki2003, Ferlaino_roton2018, Ferlaino_roton2019, Pfau_roton2021, Pfau2_roton2021}, as illustrated in Fig.~\ref{Fig_exc_spec_dipolar_gas}(a,b).

\begin{figure}
	\centering
	\includegraphics[width=\columnwidth]{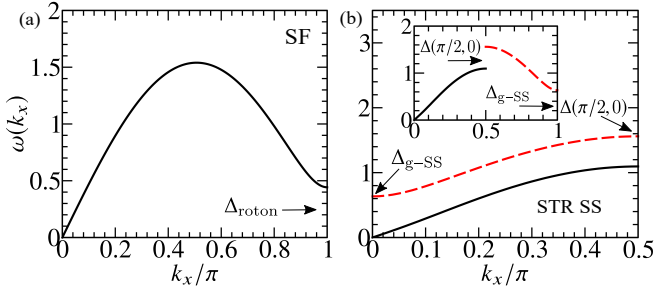}
	\caption{Excitation spectrum of hardcore bosons obtained from MF theory at zero temperature for (a) superfluid (SF) and (b) striped supersolid (STR SS) phase, in the regime $2V_{2} > V_{1}$. The Brillouin zone in (b) is folded along $k_{x}$, while the inset shows excitation within the extended scheme. Various energy gaps are denoted by the different arrows. Figures are reproduced from Ref.~\cite{MalakarJSM2023}. $\copyright$ IOP Publishing Ltd. All rights reserved.}      
\label{Fig_exc_spec_hc_bosons}
\end{figure}

In the case of lattice bosons, the collective excitations obtained from the EBHM also reveal the appearance of a roton-like mode softening at $\vec{k} = (\pi,\pi)$, or at $\vec{k}=(\pi,0)$ and $(0,\pi)$ [see Fig.~\ref{Fig_exc_spec_hc_bosons}(a)], prior to the formation of SSs with CB or STR ordering on a square lattice, respectively, depending on the interaction strengths \cite{Sinha2005, MalakarJSM2023}. 
Such roton modes have also been observed in quantum gases with cavity-mediated interactions, serving as an indication of the SS phase \cite{Esslinger_roton2012}.

The spontaneous breaking of a continuous symmetry is always associated with the appearance of a {\it Goldstone} and {\it Higgs} mode corresponding to the phase and amplitude fluctuations of the order parameter \cite{Pekker_Varma}.  
Consequently, due to the U(1) symmetry breaking associated with non-vanishing SF order, the appearance of the gapless sound excitation is a signature of the SS phase. 
This mode has been observed experimentally in the SS phase of dipolar condensate \cite{Modugno_expt_gapless_mode_SS2019, Mossman_review2019, Pfau_expt_gapless_mode_SS2019, Ferlaino_expt_gapless_mode_SS2019} as well as the spin-orbit coupled BEC \cite{Leticia_expt_2024}. 
In a trapped condensate, the low-lying discrete collective modes can be excited by applying a suitable perturbation, which can be used to characterize such phases \cite{Stringari_book}.
For example, it was recently proposed that the `scissors mode' can be used for characterizing the NCRI in dipolar SSs \cite{Stringari_scissors_mode_PRA, Stringari_scissors_mode_PRL, Adhikari_scissors}, which has also been tested experimentally \cite{Modugno_scissors_mode}, revealing direct evidence of superfluidity in the SSs.

From the analysis of EBHM, the SS phase of lattice bosons also exhibits such a gapless sound mode, as illustrated in Fig.~\ref{Fig_exc_spec_hc_bosons}(b).
As the SS transforms to a solid phase and melts to the NF phase at finite temperature, a gap opens up at $\vec{k}=0$ of the lowest energy mode, indicating the absence of superfluidity \cite{Sinha2005, MalakarJSM2023}. 
Moreover, the lowest excitation branch of the SF phase also exhibits the mode-softening phenomena as a precursor of the SS phase, as shown in Fig.~\ref{Fig_exc_spec_dipolar_gas}(a). For a continuous transition to the S, the roton gap $\Delta_{\rm roton}$ vanishes at the transition point.
The transition from the homogeneous SF to the SS phase is accompanied by a broken lattice translational symmetry, as a result of which, the Brillouin zone is reduced depending on the nature of the density ordering, and a gap opens up at its boundary. In the case of CB SS, the gap appears at $\vec{k}=(\pi/2,\pi/2)$, whereas such a gap opens at $\vec{k}=(\pi/2,0)$ or $(0,\pi/2)$ for a STR SS, depending on the  orientation of the stripes. 
For example, such gap opening associated with the STR SS has been shown in Fig.~\ref{Fig_exc_spec_dipolar_gas}(b).
Apart from the gapless modes, higher energy gapped modes in the SSs have also been obtained from the analysis of EBHM. 
Moreover, the lowest gapped mode becomes gapless at the continuous SF-SS transition boundary \cite{MalakarJSM2023}.
The transitions as well as the melting pathways of the SSs in the presence of thermal fluctuations can also be probed from the characteristic features of their excitations \cite{Malakar_triangular2020, MalakarJSM2023}. These features in the excitations are also present for the paired SS phase in the lattice \cite{Malakar2023}. Additionally, the structural transition between the different lattice-SS phases can also be probed from the changes in their respective excitation spectrum \cite{MalakarJSM2023}.

Unlike the lattice systems, the SS phases of dilute atomic gases break continuous translational symmetry.
Consequently, additional gapless Goldstone modes appear, which have been both theoretically explored \cite{Blakie_exc_spec_2024, Moroni2012, Pohl2013, Rossi2013, Blakie2023, Ferlaino_expt_gapless_mode_SS2019, 
Stringari_sound_2024, Blakie_exc_spec_2024, Blakie_exc_spec2_2024, Blakie2025, Zwerger2021, Watanabe2012, Blakie2024_binary} and experimentally observed \cite{Modugno_expt_gapless_mode_SS2019, Mossman_review2019, Ferlaino_expt_gapless_mode_SS2019}.
On the contrary, such gapless modes are absent in lattice-SSs due to the breaking of the discrete translational symmetry in the lattice.
For homogeneous systems, a distinct gapped Higgs mode can also arise in addition to the gapless modes as a consequence of the continuous translational symmetry breaking; however, this excitation can be damped \cite{Pekker_Varma}.
As depicted in Fig.~\ref{Fig_exc_spec_dipolar_gas}(c), the excitation spectrum for dipolar SS exhibits both the gapless Goldstone and the gapped Higgs modes \cite{Pfau_Higgs_mode}.
Moreover, for the SS phase of BEC coupled to two optical cavities, both kind of modes have already been monitored experimentally \cite{Esslinger_expt2_2017}.

The spontaneous breaking of both the continuous U(1) and translational symmetries in a SS can be studied from the behavior of its excitations, which are closely linked to its elastic properties and can be captured within the Hydrodynamic description \cite{Pomeau_Rica2007_1, Dorsey2010, Stringari_sound_2024, Blakie_exc_spec_2024, Blakie_exc_spec2_2024, Rakic2024, Blakie2025, Zwerger2021}. 
The appearance of multiple gapless bands, including the spin excitations in the spin-orbit coupled BEC, signifies the SS phase \cite{Stringari_superstripes2013, Han_Pu_stripes2017, Stringari_stripes2021, Stringari_stripedynamics2023}, and such modes have recently been probed in an experiment \cite{Leticia_expt_2024}.
Furthermore, in trapped dipolar gases, supersolidity has been confirmed by investigating compressional modes, which are particularly associated with the breaking of translational symmetry \cite{Modugno_expt_gapless_mode_SS2019,Mossman_review2019, Ferlaino_expt_gapless_mode_SS2019}.
Additionally, the structure of the SSs, particularly in lattice systems can also be probed from the excitation bands. 
The characteristic features of the excitation spectrum can therefore reveal both supersolidity as well as the underlying density ordering associated with it.

\section{Discussion and outlook}
\label{discussion_and_outlook}
This review presents a comprehensive overview of the theoretical and experimental progress achieved till date for exploring the intriguing SS phase, especially in ultracold atomic setups. 
While $^4$He exhibits both SS and solid phases, as well as roton mode softening, the existence of a SS phase has yet to be confirmed experimentally.
On the other hand, experiments with cold atoms have successfully demonstrated the existence of the SS phase, establishing themselves as promising platforms for exploring such exotic phases of quantum matter. 
In contrast to $^4$He, the dilute atomic gases can form almost pure condensates, where their phase-coherent properties and the ability to image density modulations serve as unambiguous indicators of supersolidity. 
In these systems, the main challenge of inducing the crystalline order is overcome by the long-range two-body interactions in dipolar condensates \cite{Lewenstein_review2009, Zoller_review2012, Ferlaino_review2023}, which have indeed led to the observation of the long-sought SS phase \cite{Bottcher_review2021, Stringari_review2023, Pfau_expt_SS2019, Modugno_expt_SS2019, Ferlaino_expt_SS2019, Ferlaino_expt_SS2021, Ferlaino_exptnature_SS2021, Ferlaino_expt_SS2022, Giamarchi_expt_SS2021}.
Similarly, the Rydberg atoms and polar molecules also hold promise for achieving the SS phase as they also possess long-range interactions \cite{Bloch_expt_Rydberg2012, Pfau_Rydberg_review2012, Lahaye_review2020, Tian_review2021, Rey_review2017, Ye_review2017, BEC_polar2024}. 
More recently, long-range interactions have been engineered by coupling BEC with cavity modes \cite{Cavity_QED_review}.
Alternatively, a different route to formation of the SS phase has been demonstrated in spin-orbit coupled BECs, exhibiting a super-stripe phase \cite{Spielman2011, Ketterle2017, Leticia_expt_2024}.
For homogeneous systems, supersolidity is always accompanied by the appearance of additional Goldstone modes corresponding to the breaking of continuous translational symmetry \cite{Modugno_expt_gapless_mode_SS2019, Mossman_review2019, Ferlaino_expt_gapless_mode_SS2019}.
The emergence of supersolidity in dipolar condensates is more delicate, as the anisotropic nature of the interactions leads to collapse, which is prevented by quantum fluctuations that stabilize the system.
Consequently, an intriguing SS phase consisting of an array of phase coherent droplets is formed for a narrow range of parameter, reflecting its fragile nature \cite{Pfau_expt_SS2019, Modugno_expt_SS2019, Ferlaino_expt_SS2019, Ferlaino_expt_SS2021, Ferlaino_exptnature_SS2021, Ferlaino_expt_SS2022, Giamarchi_expt_SS2021}.

Furthermore, the use of optical lattices opens up the possibility of realizing the lattice-SS phase, which not only stabilizes the dipolar gas but also provides an opportunity to explore the effects of correlations by tuning the hopping strength.
Although the Goldstone mode associated with density ordering is absent in lattice systems--where only discrete translational symmetry can be broken--these setups offer better insights into how SSs emerge from density wave phases through the softening of particle and hole-like excitations. Additionally, the interplay between the on-site and long-range interactions in these lattice setups can result in the formation of various types of solids \cite{Santos2002, Sun2007, Danshita2009, Sansone2014, Zhang2021, Zhang2022, Sinha2005, Pinaki2005, Demler2006, Iskin2011, Kimura2011, Kawashima2012_2}, which in turn can give rise to SSs with different structures, such as the CB and STR phases in a square lattice.
Several types of crystalline order have already been realized for dipolar atoms in optical lattices \cite{Griener2023}, where the resulting density patterns can be directly visualized using quantum gas microscopy \cite{quantum_gas_microscope}. 
For lattice-SSs, stabilizing the two coexisting orders--namely, the superfluidity and the density ordering--is crucial, as these competing orders typically exhibit a tendency toward phase separation \cite{Batrouni2000}. By lowering the on-site interaction, a stacked supersolid phase emerges with multiple occupancy of bosons at certain sites of the lattice, which enhances the stability of SSs \cite{MalakarJSM2023, Iskin2011}.
The evidence of a cluster SS has also been found for dipole-blockaded bosons using QMC simulations \cite{Cinti2010}.
In this context, it is worth mentioning that, rather than a perfect crystal with a single atom per site, a cluster of atoms is a better candidate for supersolidity, as pointed out in Ref.~\cite{Imry1975}.
From this point of view, inducing the crystalline order in homogeneous SFs is a more suitable route for achieving supersolidity. Similarly, bosons with long-range interactions in optical lattices can also host insulating DW phases with larger fillings at each site, unlike real crystals, which in turn can facilitate the formation of SSs by melting such DW phases through tuning the hopping strength.
The remarkable progress in cavity QED setups enables the generation of cavity-mediated long-range interactions, which lead to the formation of SSs of cold atoms loaded in an optical lattice coupled to a cavity mode, setting the stage for exploring novel quantum phases.
 
Apart from cold atoms, the edges of the quantum hall sample have been predicted to exhibit chiral SS behavior \cite{quantum_hall_SS}. 
Excitons in semiconductors also have the potential to host supersolidity \cite{Peeters_2023, Littlewood2011, Kavokin2012, Pfieffer2024}.
Recently, the evidence of a magnetic analogue of supersolidity (spin SS) has also been explored in synthesized triangular anti-ferromagnetic materials \cite{spin_SS_material, Zheludev2024}. 
Spatially modulated SFs have also been realized experimentally in layered $^4$He films adsorbed on graphite substrate \cite{Choi2021,Rao2021}, although characterizing it as SS is still questionable since the spontaneous breaking of the translational symmetry is absent.
Furthermore, the quantum engineering techniques allow the possibility of generating fascinating structures in SSs. 
Topological features like vortices have already been demonstrated in dipolar SSs \cite{vortices_SS_expt, vortices_Stringari2020, vortices_Reatto2021}, confirming their SF nature. 
Interestingly, such rotating SSs exhibit `glitches' similar to neutron stars \cite{glitches_rotating_SS, pulsar_glitches}, which opens a new pathway for quantum simulation of stellar objects. 
It has also been predicted that the inner crust of neutron stars can exhibit SS behavior \cite{pasta_phase_neutron_star}.
Moreover, ring-shaped SSs are also expected to emerge in toroidal traps \cite{Reimann2021, Reimann2023} and other geometrical confinements \cite{Boronat2024, Roccuzzo2022}.
Even a chiral SS phase can be created in a synthetic spin-orbit coupled Rydberg dressed BEC \cite{chiral_SS_SOC1, chiral_SS_SOC2}. 
A recent proposal to engineer pair hopping of bosons in an optical lattice \cite{Ripoll2008, Zhang2009} opens up the possibility to study the exotic paired SF and SS phases \cite{Hauke2012, Sandvik2019, Wang2013, Wang2013_2, Malakar2023}. In this context, two-component bosons and bilayer-optical lattices are also promising candidates \cite{Kuklov2004, Hu2009, Menotti2010, Menotti2022}.

Beyond its creation, exploring the structural transitions of SSs is another important direction for research.
For instance, the honeycomb SS of dipolar gas has been predicted which exhibits the emergence of Dirac points similar to Graphene, as well as a structural transition following a quench \cite{Blakie2025}.
In addition, the fate of the SS phases at finite temperatures is also a pertinent issue. Due to the coexistence of competing orders, there can be different melting pathways where one of the orders vanishes before the other. Moreover, the transient nature of SS and its creation through thermal quenches have already been explored in dipolar gases \cite{Ferlaino_expt_SS2019, Ferlaino_expt_SS2021, Ferlaino_expt_SS2022, Ferlaino_nat_comm2023}. 
Observing the SS phases as metastable states is also a related issue \cite{Modugno_expt_SS2019, Zhu2023, Zhang2024, Blakie2025}. 
Furthermore, the observation of supersolidity in a driven dissipative polariton condensate in a photonic crystal waveguide \cite{nature_SS_photons,PRL_SS_photons} also motivates the search for the SS phase as a fascinating non-equilibrium state in other open quantum systems, such as Rydberg and cavity QED setups with inherent loss processes \cite{Fazio_solidphoton2015, Ritsch2018, Hofstetter2019_non_eqSS, Ray2024}.

While the quest for supersolidity began with $^4$He, ultracold atomic setups have emerged after almost 70 years as an ideal platform for realizing such exotic phases of synthetic matter, paving the way to explore a diverse range of SS phases and their intriguing properties.

\section*{Acknowledgment}
We would like to thank our collaborators, Sayak Ray, Manali Malakar, Dilip Angom, Krishnendu Sengupta, and D. Kovrizhin. This article is dedicated to the memory of our collaborator, G. V. Pai.

\appendix

\section{Mean-field description}
\label{appendix1}
Within the mean-field (MF) prescription, the correlations between the different lattice sites (or between the clusters) can be neglected, so that the wavefunction can be written as a product state $\ket{\psi} = \prod_{i} \otimes  \ket{\psi_{i}}$. Using the Gutzwiller approach, the wavefunction at each site can be written as $\ket{\psi_{i}}=\sum_{n}f^{(i)}_{n}\ket{n}$ \cite{Krauth1992}, where $f^{(i)}_{n}$ denote the Gutzwiller amplitudes and $\ket{n}$ represents the number states. By minimising $\bra{\psi}\hat{\mathcal{H}}-\mu\hat{N}\ket{\psi}$, subjected to the constraint $\sum_{n}|f^{(i)}_{n}|^2=1$, the steady state amplitudes $\bar{f}^{(i)}_{n}$ are obtained which describe the ground state. 
For dynamics, a straight-forward extension of the Gutzwiller method can be achieved by introducing the time-dependent amplitudes $f^{(i)}_{n}(t)$.
Minimizing the action $S=\int dt \bra{\psi}\dot{\iota}\partial/\partial t-(\hat{\mathcal{H}}-\mu\hat{N})\ket{\psi}$ yields the corresponding equations of motion,
\begin{eqnarray}
\dot{\iota}\frac{\partial}{\partial t}f^{(i)}_{n} &=& \left[\frac{U}{2}n(n-1)-\mu+\tilde{V}_{i}n-\lambda_{i}\right]f^{(i)}_{n} \notag\\
&&-t\left(\phi^{*}_{i}\sqrt{n+1}f^{(i)}_{n+1}+\phi_{i}\sqrt{n}f^{(i)}_{n-1}\right)
\end{eqnarray}
where $\tilde{V}_{i} = \sum_{j\neq i}n|f^{(j)}_{n}|^2$, $\phi_{i}=\sum^{\rm NN}_{j\neq i}n|f^{(j)}_{n}|^2$, and $\lambda_{i}$ denote the Lagrange multiplier for each lattice site, which can be obtained from the steady state values $\bar{f}^{(i)}_{n}$.
The linearization of the above equations around the ground state considering the time evolution of a small perturbation 
$\delta f^{(i)}_{n}(t) =e^{\dot{\iota}\omega t} \delta f^{(i)}_{n} (0)$, 
\begin{eqnarray}
f^{(i)}_{n}(t) =  \bar{f}^{(i)}_{n} + \delta f^{(i)}_{n}(t),
\end{eqnarray}
gives the collective excitation frequency $\omega$ \cite{Sinha2005, Dima2007}. 
This method has also been applied to the dissipative Rydberg systems \cite{Sayak_Rydberg2016}.

The MF method can easily be generalized to finite temperatures, where the density matrix assumes the product form $\hat{\rho}=\prod\hat{\rho}_{i}$. By constructing the mean-field Hamiltonian $\hat{\mathcal{H}}^{\rm MF}_{i}$ at each lattice site (or cluster), the corresponding equilibrium density matrix at finite temperature $T=1/\beta$, $\bar{\hat{\rho}}=\prod \bar{\hat{\rho}}_{i}$ can be determined. The local order parameters, such as $\langle \hat{n}_{i} \rangle={\rm Tr}(\bar{\hat{\rho}}_{i}\hat{n}_{i})$ and $\langle \hat{a}_{i} \rangle={\rm Tr}(\bar{\hat{\rho}}_{i}\hat{a}_{i})$ can then be obtained from $\bar{\hat{\rho}}_{i}=e^{-\beta\mathcal{H}^{\rm MF}_{i}}/Z_{i}$ in a self-consistent manner, where $Z_{i}={\rm Tr}(e^{-\beta \mathcal{H}^{\rm MF}_{i}})$. 
The effect of correlations can also be introduced by using the cluster mean-field (CMF) method \cite{Danshita2012, Danshita_triangular2012,  Abouie2017, Suthar2020, Suthar2022, Ray2022, Malakar_triangular2020, MalakarJSM2023, Malakar2023, Abouie2020, Ray2024} and other techniques \cite{Menotti2022}.

A nearly pure condensate of dilute gases can be described by a macroscopic wavefunction or a classical field $\psi(\vec{r})$, satisfying the non-linear Gross-Pitaevskii equations (GPE),
\begin{eqnarray}
\dot{\iota}\hbar\frac{\partial }{\partial t}\psi(\vec{r},t) &=& \Big[ -\frac{\hbar^2}{2m}\nabla^2+V_{\rm ext}(\vec{r})+\!\!\int\!\! d\vec{r}' U(\vec{r}-\vec{r}')|\psi(\vec{r}',t)|^2 \notag\\
&&+ \Delta \mu_{\text{\tiny LHY}}[\psi(\vec{r},t)]\,\Big]\,\psi(\vec{r},t),
\end{eqnarray}
where $V_{\rm ext}(\vec{r})$ is the external potential and $U(\vec{r}-\vec{r}')$ is a two-body interaction potential. For dipolar interaction, the two-body potential takes the following form,
\begin{eqnarray}
U(\vec{r}-\vec{r}') = g\delta(\vec{r}-\vec{r}') + \frac{3g_{dd}}{4\pi}\frac{(1-3\cos^2{\theta})}{|\vec{r}-\vec{r}'|^3},
\end{eqnarray}
where the first part denotes the short-range contact interaction with strength $g=4\pi\hbar^2a_{s}/m$ determined by the $s$-wave scattering length $a_{s}$. On the other hand, the second part represents the long-range interaction with strength $g_{dd}=4\pi\hbar^2a_{dd}/m$, where $a_{dd}=m\mu_{0}\mu^2_{m}/12\pi\hbar^2$ is the dipolar length determined by the magnetic moment $\mu_{m}$ of the particles, and $\theta$ is the orientation of the dipoles along the $z$-axis.
By linearizing the GPE around the equilibrium phase, the collective modes can be extracted, which are the same as the excitations obtained from the Bogoliubov approach \cite{Stringari_book}. It should be noted that, for the formation of dipolar SSs, the inclusion of the beyond MF term $\Delta\mu_{\text{\tiny LHY}}[\psi(\vec{r})]=\gamma |\psi(\vec{r})|^3$ in the GPE becomes essential, where $\gamma = \frac{128\pi\hbar^2}{3m}a_{s}\sqrt{a^3_{s}/\pi}{Q}_{5}(a_{dd}/a_{s})$ with $Q_{5}(x) = {\rm Re}\{\int^{1}_{0}\! du [1+x(3u^2-1)]^{5/2}\} $ (see for example Ref.~\cite{Blakie_exc_spec_2024, Fischer2006, Pelster_fluctuation}).

\end{document}